\documentclass[final,5p,times,twocolumn]{elsarticle}

\usepackage{graphicx}
\usepackage[update,prepend]{epstopdf}
\usepackage{amsmath}

\usepackage{amssymb}
\usepackage{lineno}

\usepackage{xspace}
\journal{Nucl. Instrum. Methods Phys. Res., Sect. A}

\begin{document}
\begin{frontmatter}



\title{Absolute Position Measurement in a Gas Time Projection Chamber via Transverse Diffusion of Drift Charge}

\author{P.~M.~Lewis}
\author[]{S.~E.~Vahsen\corref{cor1}}
\cortext[cor1]{Corresponding author. Tel.: +1 808 956 2985.}
\ead{sevahsen@hawaii.edu}
\author{M.~T.~Hedges}
\author{I.~Jaegle} 
\author{I.~S.~Seong}
\author{T.~N.~Thorpe}
\address{University of Hawaii, 2505 Correa Road, Honolulu, HI 96822, USA}

%A concise and factual abstract is required. The abstract should state briefly the purpose of the research, the principal results and major conclusions. An abstract is often presented separately from the article, so it st be able to stand alone. For this reason, References should be avoided, but if essential, then cite the author(s) and year(s). Also, non-standard or uncommon abbreviations should be avoided, but if essential they must be defined at their first mention in the abstract itself.

\begin{abstract}
Time Projection Chambers (TPCs) with charge readout via micro pattern gaseous detectors can provide detailed measurements of charge density distributions. We here report on measurements of alpha particle tracks, using a TPC where the drift charge is amplified with Gas Electron Multipliers and detected with a pixel ASIC. We find that by measuring the 3-D topology of drift charge and fitting for its transverse diffusion, we obtain the absolute position of tracks in the drift direction. For example, we obtain a precision of 1~cm for 1~cm-long alpha track segments. To our knowledge this is the first demonstration of such a measurement in a gas TPC. This technique has several attractive features: it does not require knowledge of the initial specific ionization, is robust against bias from diffuse charge below detection threshold, and is also robust against high charge densities that saturate the detector response.
\end{abstract}


\end{frontmatter}

%\linenumbers

\section{Introduction}
\begin{figure} [h]
\centering
\includegraphics[width=8.4cm]{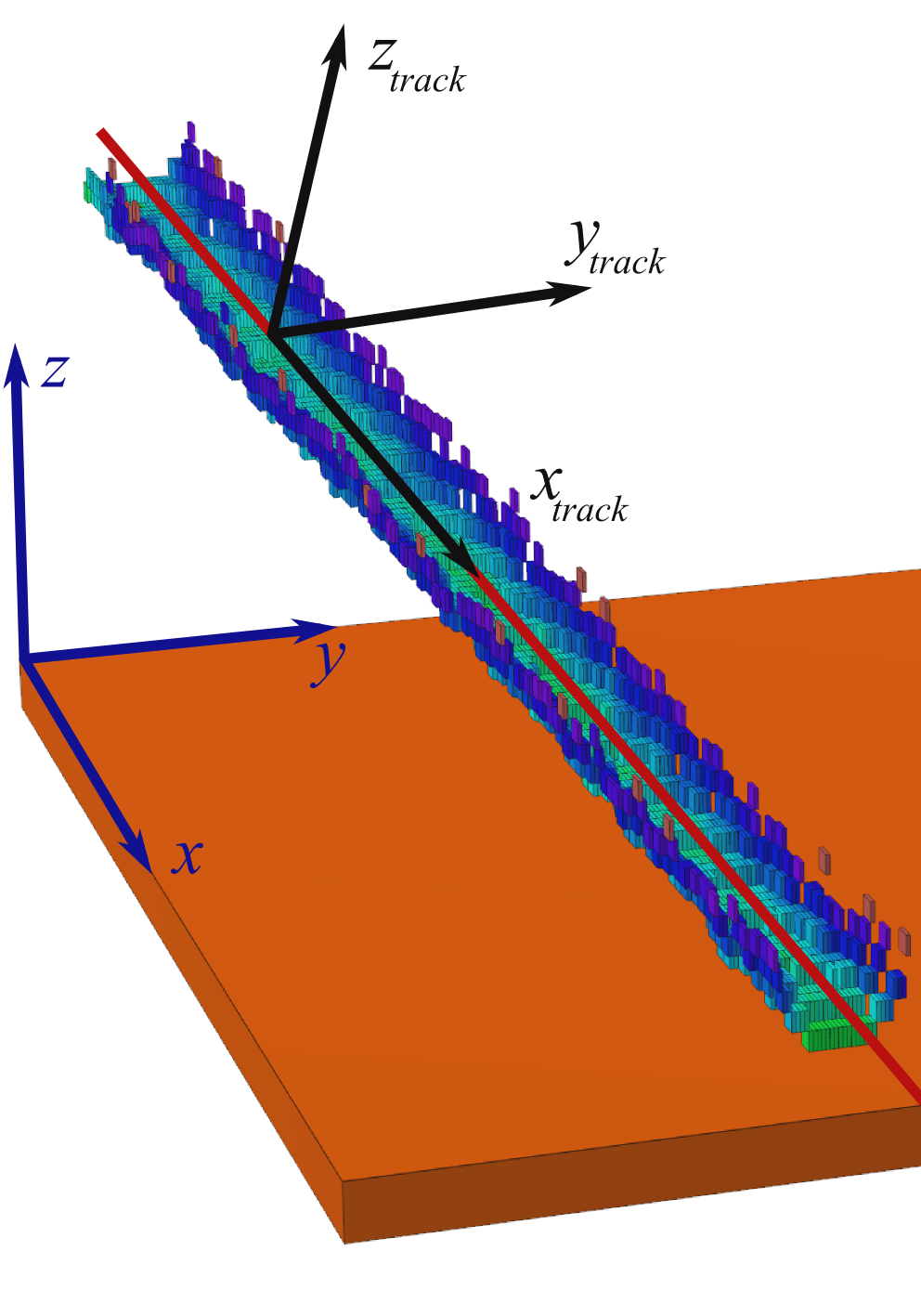}
\setlength{\abovecaptionskip}{0pt}
\caption{A typical alpha track from the dataset used in this analysis. Each box is located in 3-D space according to the $x$, $y$ and $z$ coordinates of the first threshold-crossing for the pixel under it. ToT is encoded by color, with higher ToT (lighter) in the bottom of the barrel and low ToT (darker) at the high edges. Both coordinate systems used in this paper are drawn: (blue) the chip-based Cartesian coordinate system with $z$ pointing anti-parallel to the drift direction and (black) the track coordinate system in which $y_{track}$ is the lateral displacement of a hit from the best-fit line (red) to the track shell. The track angles $\theta$ and $\phi$ are the polar angle with respect to the $z$ axis, and the azimuthal angle between the track's $x$-$y$ projection and the $x$ axis, respectively. These angles define the rotations between the two coordinate systems. The absolute origin of the track coordinate system along the track fit line is arbitrary. The bottom of the sensitive volume is shown as an orange plane.} 
\setlength{\belowcaptionskip}{0pt}
\label{fig:coordinates}
\end{figure}

Since their introduction in the 1970s, Time Projection Chambers (TPCs)~\cite{Nygren:1978rx} have found a wide range of applications in particle physics and beyond. Similar to bubble chambers, TPCs allow 3-D reconstruction of ionization distributions in large volumes. Traditionally, one limitation of the TPC has been that unless the time when ionization occurred is known, the detector only measures relative, not absolute position in the drift direction, $z$. Recently, however, Snowden-Ifft of the DRIFT collaboration demonstrated the first absolute position measurement in a negative ion drift gas TPC by using a gas mixture with multiple charge carriers~\cite{Snowden-Ifft:2014taa}. In prior work, the DRIFT collaboration had used longitudinal diffusion to estimate $z$ and veto backgrounds~\cite{Daw:2010ud}.
 However, the resolution was limited, so that the veto severely truncated the sensitive volume. Highlighting the importance of measuring $z$,  the improved technique with multiple charge carriers increased the DRIFT fiducial region from 5-10\% to 75\% of the detector volume, and reduced their background rate by at least two orders of magnitude, allowing background free operation for the first time~\cite{Dan}.

Here we show how a gas TPC with a high-density pixel readout plane is capable of measuring the absolute $z$ position of tracks with a different method, by measuring the detailed ionization density distribution of the diffused track after drift. As a bonus, the charge density distribution also provides sufficient information to measure the initial specific ionization of the track, automatically correcting for highly diffuse charge that goes undetected due to the threshold of the detector, and for high charge densities outside the dynamic range of the detector. The ionization measurement will be published separately, and is not explored further in this work.
 
\section{Detector and principle of operation}
We use a prototype directional fast-neutron detector that is an evolution of the miniature TPC described in Ref.~\cite{Vahsen:2014fba}. Details on the detector used here will be published separately~\cite{beasttpc}. In brief, the detector has a sensitive volume of $2.0\times1.68\times15$~cm$^3$ in $x \times y \times z$, filled with a 70:30 mixture of He:CO$_2$ gas at atmospheric pressure. A field cage creates a uniform drift field of $530$~V/cm in the sensitive volume. Ionization deposited in the sensitive volume drifts through a distance $z$ then is amplified with two Gas Electron Multipliers~\cite{Sauli:1997qp}, and finally is collected with an ATLAS FE-I4B pixel ASIC \cite{Aad:2008zz} (or ``chip''), which digitizes the charge signal. The ATLAS FE-I4B was developed for the recent upgrade of the ATLAS Pixel Detector, a vertex detector used in the first run of the CERN Large Hadron Collider. 

We use the chip to collect drifting charge, with high resolution in space and time. Charge is measured in units of time over threshold (ToT), which is not linear in collected charge and is based on a $40$~MHz clock. Spatially, the chip plane is segmented into $26,880$ pixels of $50\times250~\mu\text{m}^2$, arranged in a rectangular grid of 336 rows and 80 columns, dense enough to resolve the $\sim1$~mm typical width of drifted electron clouds in our TPC. We define the coordinate system so that the chip occupies the $x$-$y$ plane, perpendicular to the electric field ($+z$) and the electron drift direction ($-z$). Time bins are $25$~ns, which translates into a $z$ quantization of $250~\mu$m with our drift field. 

During normal operation of our TPC we keep three Polonium-210 alpha sources mounted on the outside of the field cage at $z=3.6$~cm, $7.6$~cm, and $11.6$~cm. We use these in-situ sources to calibrate and monitor the detector in a number of ways in real-time. We discovered that these sources can also be used to calibrate the absolute $z$ scale of the TPC, a technique we demonstrate here. 

\begin{figure}[h!]
\centering
\includegraphics[width=8.4cm]{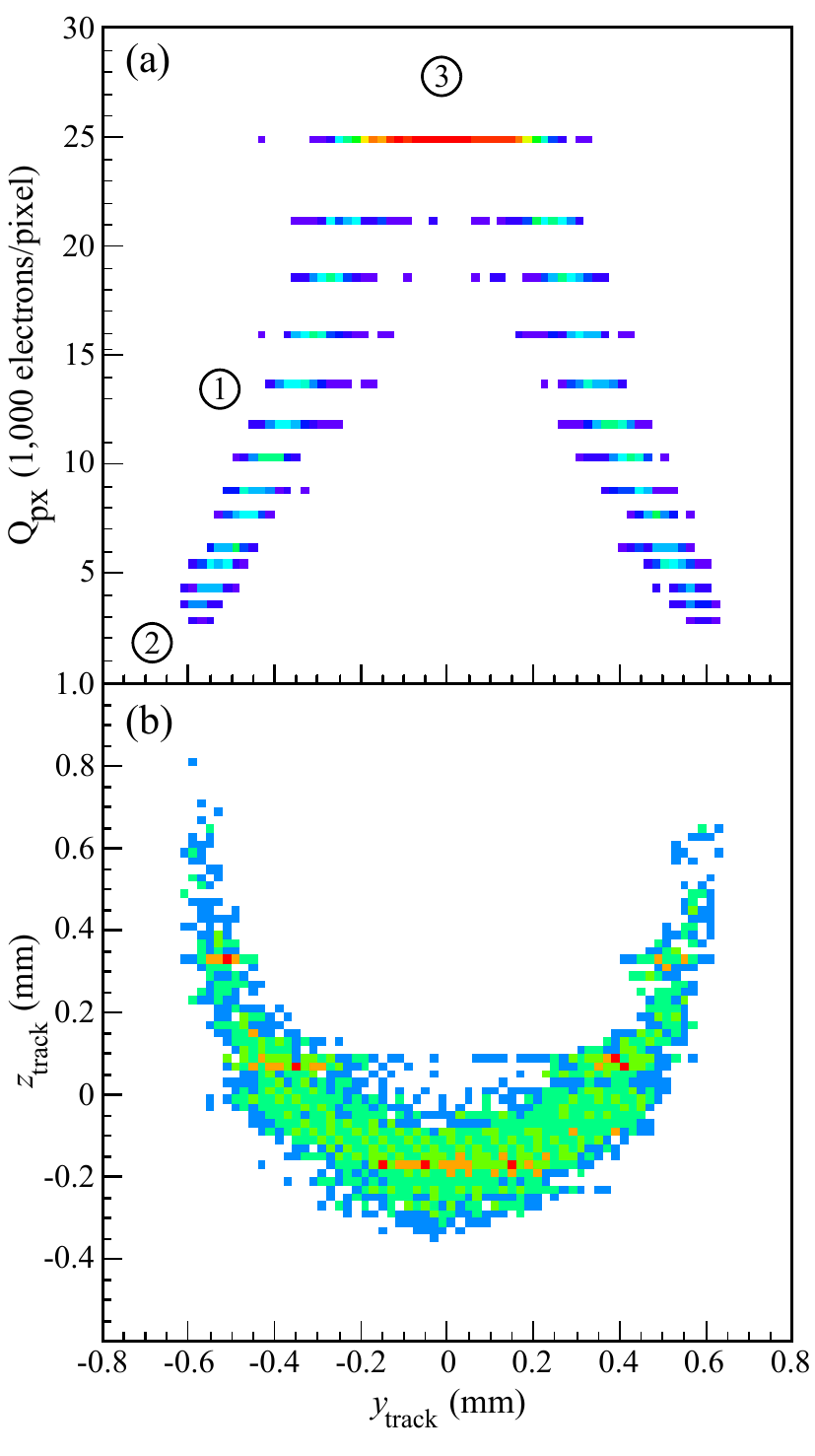}
\setlength{\abovecaptionskip}{0pt}
\caption{ Corrected pixel charge ($Q_{px}$) profile (a) and threshold contour shell coordinates (b) for a single horizontal track from the bottom alpha source. Label 1 of the profile plot (a) corresponds to the Gaussian, label 2 to the threshold, and label 3 to the saturation regions of the profile. These plots are two-dimensional histograms where the counts per bin are encoded by color: blue for the lowest count number, red for the highest.}
\setlength{\belowcaptionskip}{0pt}
\label{fig:profile_shell}
\end{figure}

\subsection{Charge cloud detection}\label{sec:detection}
As a drifting charge cloud arrives at a pixel, the 4-bit ToT counter starts once the accumulated charge reaches threshold and continues to count for as long as the pixel is over threshold. The ToT then represents the integral over time (and thus $z$) of the charge collected by each pixel. The chip output data then contain no charge structure in $z$, only the relative threshold-crossing time of the charge cloud from pixel to pixel in units of $25$~ns. Some charge is lost, such as below-threshold charge at the edge of a track or charge that arrives after a pixel has fallen below threshold. The chip therefore provides four dimensions of track topology information for each pixel, as visualized in Fig.~\ref{fig:coordinates}, namely: $x$, $y$, and $z$ spatial coordinates of the first threshold crossing for each pixel, plus ToT. The spatial coordinates define the ``shell,'' the bottom half of a cylinder which can be thought of as a single charge contour of the track. 

In our operation mode, the chip issues a trigger to the readout electronics when a pixel crosses threshold. The readout board then retrieves pixel data from pixels which crossed threshold during a fixed time window for a fixed time offset from the receipt of the trigger. Consequently, the time or $z$ axis of the pixel data has no consistent zero point --- we cannot determine the absolute $z$ of the track. In this analysis we recover absolute $z$ using distributions of the four track topology coordinates $x$, $y$, relative $z$ and ToT.

\subsection{Charge definition}
We use the chip's built-in capacitors to inject known charges into each pixel to measure an empirical correspondence between number of electrons and ToT, $N_e(\text{ToT})$. However, $N_e$ is not a consistent measure of charge density along a track. For an inclined track, the density will appear higher as the track is foreshortened on the pixel plane by a factor of $\sin\theta$ (where a horizontal track has $\theta=90^{\circ}$). We define a new quantity, the corrected pixel charge $Q_{px}$, which includes a geometric correction and can be thought of as the expected charge collected in a pixel (in electrons) if the track was horizontal:
\begin{equation}
Q_{px}(\text{ToT}) = N_e(\text{ToT})~\sin\theta.
\end{equation}\label{eqn:charge}
The charge density along the track is then $Q_{px}$ divided by the sensitive area of a pixel. This can be converted to initial ionization energy if the work function of the gas, the fraction of the chip surface that is sensitive to charge, and total effective gain are known, a step that is not necessary in this analysis. 

\section{Physics model}\label{sec:physics}
We use a simplified model of charge cloud dynamics to derive the expected variation of track topology with $z$. Specifically, we consider a line of ionization charge with an initial delta-function charge cross-section. The constituent charges diffuse in space in a 3-D random walk over time, giving a charge profile which is Gaussian in cross-section with a constant integral and a width that is proportional to the square root of time, $\sigma_z(t)\propto \sqrt{t}$. In the presence of a linear potential, the charge line drifts to the detector plane at a constant velocity, implying that the width of the charge distribution is also proportional to the square root of absolute $z$, or $\sigma_z(z)=B \sqrt{z}$. There is a fixed $z$-independent width $A$ associated with the resolution of the readout plane, leading to an expected relationship between the width of the charge distribution and absolute $z$ given by the quadrature sum of both terms~\cite{Vahsen:2014fba}:
\begin{equation}
\sigma(z) = \sqrt{A^2 + B^2  z}.
\label{eqn:sig_z}
\end{equation}
\noindent
Our model does not include two minor effects: first, re-absorption of drift electrons with gas impurities such as oxygen, which results in a linear decrease in the total charge of the track over drift distance. Second, based on the track position along the Bragg curve, we expect small changes in the initial ionization density along the track ($dE/dx$), a $z$-independent effect. We confirm that neither effect impacts our measurement of absolute $z$.

\begin{figure} [h]
\centering
\includegraphics[width=8.4cm]{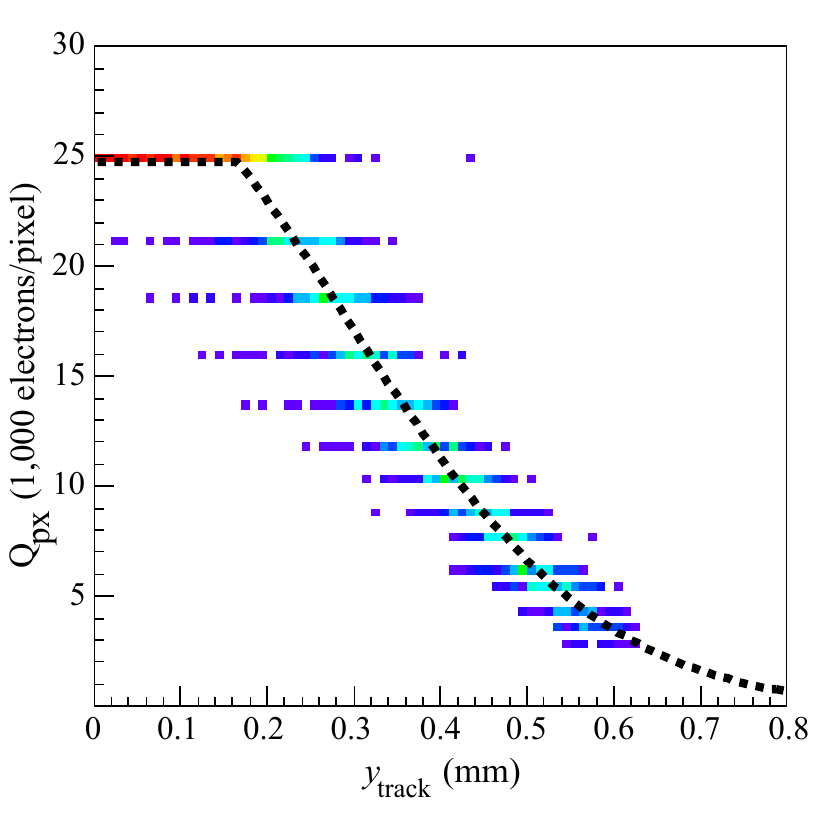}
\setlength{\abovecaptionskip}{0pt}
\caption{ Fit to the folded version of the corrected pixel charge ($Q_{px}$) profile shown in Fig.~\ref{fig:profile_shell} using Eqn.~\ref{eqn:fitfcn} as the fit function.} 
\setlength{\belowcaptionskip}{0pt}
\label{fig:profile_fit}
\end{figure}

\section{Data and analysis}\label{sec:analysis}
Our goal is to demonstrate a measurement of absolute $z$ using the charge cloud topology information provided by the chip. Specifically, we fit a Gaussian function to the corrected pixel charge $Q_{px}$ versus lateral distance from track center (the charge ``profile"). If we know the parameters $A$ and $B$ in Eqn.~\ref{eqn:sig_z} then the fitted Gaussian width $\sigma$ would uniquely measure absolute $z$ independent from initial track ionization energy. To this end, we focus on a precision determination of the diffusion parameters $A$ and $B$. 

Our first task is to verify the model using the embedded alpha sources, then measure the parameters in Eqn.~\ref{eqn:sig_z}, and finally test the performance of the absolute $z$ measurement. 

\begin{figure*}[h!t]
\centering
\includegraphics[width=16cm]{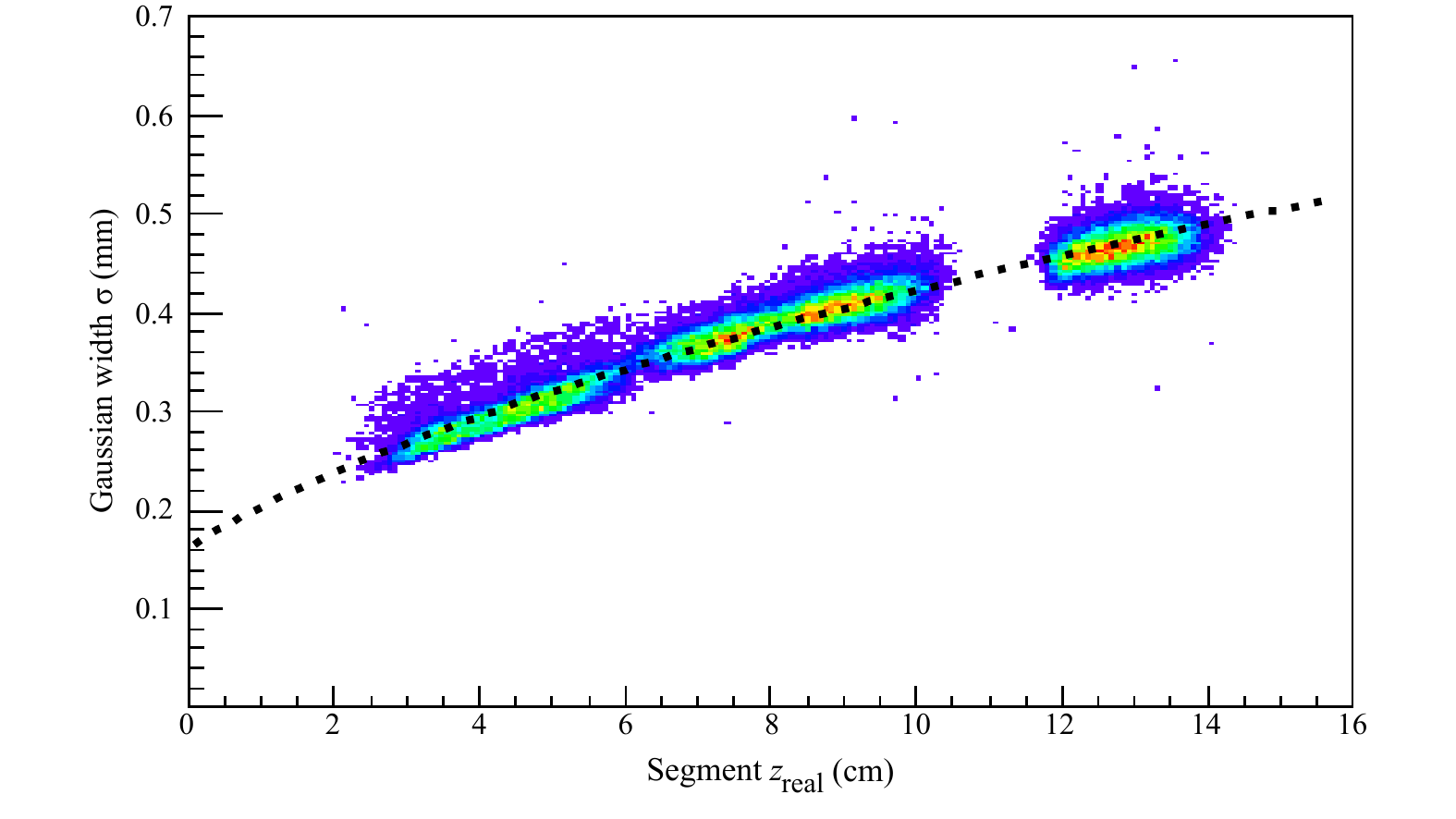}
\setlength{\abovecaptionskip}{0pt}
\caption{Result of fitting the diffusion equation, Eqn.~\ref{eqn:sig_z}, to the distribution of fitted charge profile Gaussian widths versus $z_{\text{real}}$, for $1$~cm track segments.}
\setlength{\belowcaptionskip}{0pt}
\label{fig:sig_v_z}
\end{figure*}

\subsection{Data}\label{sec:data}
Alpha tracks from the in-situ sources provide an ideal calibration and test of the absolute $z$ determination. We use $100,000$ alpha tracks generated during a run with no external sources. We fit these tracks with a linear unweighted least-squares fit to the charge shell and obtain the track polar and azimuthal angles, $\theta$ and $\phi$. Using these track angles and the known position of each alpha source in the detector, we can assign a coordinate in 3-D to each pixel hit in a track; crucially, this allows us to know the ``real'' $z$ position of each hit, which we label $z_{real}$. 

For each track we rotate to a new ``track'' coordinate system to look ``down the barrel" of the track; the track now lies along the $x_{track}$ axis, the U-shaped shell opens towards the $z_{track}$ direction, and the lateral track width dimension is now $y_{track}$ (see Fig.~\ref{fig:coordinates}). This coordinate system facilitates easy viewing of the track charge profile and shell (Fig.~\ref{fig:profile_shell}). In order to localize $z$-dependent effects, we separate the last $2$~cm of each track into two $1$~cm ``segments" which we evaluate independently for the remainder of the analysis. 

The alpha segment data constitutes both calibration and validation data. We use the segments to generate a distribution of cloud shape vs. $z_{real}$, then extract a functional approximation of that distribution to assign absolute $z$ based on cloud shape. The spread of the distribution around the functional approximation is a measure of the accuracy of the approximation and thus the absolute $z$ measurement technique. 

\subsection{Building charge profiles}\label{sec:profiles}
In the track coordinate system, that is, looking down the axis of the track, the charge profile has three main features, labeled in Fig.~\ref{fig:profile_shell} (a) for reference. Feature $1$ is the predicted Gaussian charge cross-section. Feature $2$ is a cutoff in the Gaussian tail at a certain charge, due to the chip's threshold setting, here tuned to a level of~$\sim3,000$ electrons per pixel. Feature 3 is a flat saturation region corresponding to the high end of the ToT scale, tuned to $\sim25,000$ electrons. Saturation primarily occurs for highly inclined ($\theta$ not near $90^{\circ}$) or low-$z$ tracks. The threshold and saturation charges are given by $Q_{px}(\text{ToT}_{min})$ and $Q_{px}(\text{ToT}_{max})$ and are therefore dependent on the track inclination $\theta$. As shown in Fig.~\ref{fig:profile_shell}, even $> 2$~cm-long flat alpha tracks have clean charge profiles due to the slow dependence of ionization energy density on distance from source.

\subsection{Profile fitting}\label{sec:fitting}
Prior to fitting we fold the profile distribution about the $z_{\text{flat}}$ axis. The physics-motivated fitting function is given by:
\begin{equation}
f(y_{\text{track}}) = min\left (  Q_{px}(\text{ToT}_{max}),h \cdot \text{Exp}\left[ -\frac{y_{\text{track}}^2}{2\sigma^2} \right] \right ),
\label{eqn:fitfcn}
\end{equation}
\noindent
that is, a Gaussian of width $\sigma$ and maximum height $h$ centered at $y_{track}=0$ or the saturation charge $Q_{px}(\text{ToT}_{max})$, whichever is smallest. We use a least-squares fit of this function to the 2D scatterplot of points in the $Q_{px}$ profile. An example of this fit applied to the charge profile of a flat track ($\theta=90^{\circ}$) is shown in Fig.~\ref{fig:profile_fit}. 

For highly saturated segments, such as those from steeply angled tracks near $z=0$, the Gaussian parameters are poorly constrained, in particular the height $h$. Consequently, for the determination of the diffusion parameters $A$ and $B$ we select only fits with a normalized chi-squared below $1.5$ (efficiency $0.55$).

\section{Results}\label{sec:results}
Our primary result is a least-squares fit, shown in Fig.~\ref{fig:sig_v_z}, matching Eqn.~\ref{eqn:sig_z} to the scatterplot of Gaussian widths versus known $z$ for $1$~cm segments selected from $100,000$ alpha tracks. The good quality of this fit validates the physical model. The fit parameters obtained are $A=180\pm20~\mu$m and $B=123\pm7~\mu$m. The former is the effective point resolution of the readout plane. The latter is the diffusion per $\sqrt cm$, consistent with the Magboltz prediction~\cite{magboltz}, further validating the model. 

Now we can use these diffusion parameters to predict absolute $z$ for any segment of ionization charge detected:
\begin{equation}
z = \frac{\sigma^2-A^2}{B^2}.
\end{equation}\label{eqn:absz}
The absolute $z$ accuracy is the horizontal deviation of the distribution from the fit curve in Fig.~\ref{fig:sig_v_z}. For all segments, without any selections on fit quality, the absolute $z$ measurement is $1-2$~cm throughout the full $z$ range. Of these segments, a portion have a normalized chi-squared under $1.5$. For these high-quality fits the absolute $z$ precision is $0.6-1.0$~cm. The chi-squared selection has an efficiency of $0.55$ overall: $0.34$, $0.51$ and $0.95$ for the bottom, middle and top sources, respectively. For nuclear recoil tracks, such as neutron or WIMP recoils, the absolute $z$ precision will depend strongly on the length of the track. We have already successfully applied the technique described here to mm-length alpha track segments, and plan to study nuclear recoils next.

\section{Additional results and techniques}\label{sec:additional}
The techniques developed for this analysis can be applied to a number of related questions which are largely beyond the scope of this paper. We summarize them here for completeness. 

The first extension of this technique is a recovery of charge loss due to threshold and saturation effects. We consider a Gaussian which is defined by the shape parameters determined by a charge profile fit. The Gaussian extends beyond the threshold region and above the saturation region, therefore it represents the entire charge profile without losses. A full discussion of this technique is not possible here, but we have observed that recovery of charge lost to saturation is possible even for very saturated tracks. The recovery of the full charge for each segment also allows for a determination of the dependence of ionization energy on the distance from the source. We find slopes consistent with zero, suggesting that $dE/dx$ can be considered uniform for the sources within the TPC. 

As a complementary approach to the charge profile fit method, we also attempted to calibrate absolute $z$ using fits to the shape of the charge shell. The physics model is somewhat more complicated than in the profile case due to the interplay between re-absorption, diffusion and chip threshold. We fit ellipses to the shell scatterplots $z_{track}$ vs. $y_{track}$ for the track segments and determine that no $1:1$ correspondence between shell semi-major or semi-minor widths and $z_{real}$ is likely. However, there is a relationship between the shell fit residuals and $z$ that is apparently linear in $z$. Absolute-$z$ determination with this method is only accurate to $~4$~cm and therefore is not included in this analysis.

\section{Summary and discussion}
We have presented a method for measuring absolute $z$ in TPCs using charge cloud topology information. The key technical features that enable this technique are high spatial resolution of the pixel chip and charge measurement in each pixel. While high spatial resolution alone is enough to measure the diffusion width of tracks, this width also depends strongly on nuisance parameters, such as the detector threshold, gas purity, and the specific ionization of the track. By also utilizing charge information, the technique presented here does not require knowledge of the initial ionization energy and is tolerant of saturation and threshold effects. 

It is possible to further optimize the absolute $z$ precision using a lower threshold and higher maximum ToT scale, though gains may be small except for very low $z$. This technique is best used as a live calibration of the absolute $z$ scale since the diffusion parameters may depend on gas quality, gain, drift velocity and other factors and thus may not be stable over time. Embedding one or more sources such as Polonium-210 in the vessel during operation can provide time-dependent calibrations.

The technique used here also provides a way to recover charge lost due to chip threshold and saturation effects, which naturally provides a measurement of gas impurities by way of charge loss over drift distance. This will be reported separately.

\section{Acknowledgements}
We thank Jared Yamaoka and Steven Ross, former Hawaii group members who designed a pixel readout board used in this work. We thank John Kadyk and Maurice Garcia-Sciveres of Lawrence Berkeley National Laboratory for continued support and collaboration, and Kamalu Beamer for assistance during detector construction. We acknowledge support from the U.S. Department of Homeland Security under Award Number 2011-DN-077-ARI050-03 and the U.S. Department of Energy under Award Number DE-SC0007852.

\section{References}
\bibliographystyle{elsarticle-num}

\begin{thebibliography}{00}
\bibitem{Nygren:1978rx} 
  D.~R.~Nygren and J.~N.~Marx,
  ``The Time Projection Chamber,''
  Phys.\ Today {\bf 31N10}, 46 (1978).
  %%CITATION = PHTOA,31N10,46;%%
  %18 citations counted in INSPIRE as of 05 Oct 2013

\bibitem{Snowden-Ifft:2014taa} 
  D.~P.~Snowden-Ifft,
  ``Discovery of multiple, ionization-created CS2 anions and a new mode of operation for drift chambers,''
  Rev.\ Sci.\ Instrum.\  {\bf 85}, 013303 (2014).

\bibitem{Daw:2010ud} 
  E.~Daw, J.~R.~Fox, J.~L.~Gauvreau, C.~Ghag, L.~J.~Harmon, M.~Gold, E.~R.~Lee and D.~Loomba {\it et al.},
  ``Spin-Dependent Limits from the DRIFT-IId Directional Dark Matter Detector,''
  Astropart.\ Phys.\  {\bf 35}, 397 (2012)
  [arXiv:1010.3027 [astro-ph.CO]].
  %%CITATION = ARXIV:1010.3027;%%
  %30 citations counted in INSPIRE as of 04 Oct 2014

\bibitem{Dan}
 D.~P.~Snowden-Ifft, {\it personal communication}, October 2014

%\cite{Vahsen:2014fba}
\bibitem{Vahsen:2014fba} 
  S.~E.~Vahsen, M.~T.~Hedges, I.~Jaegle, S.~J.~Ross, I.~S.~Seong, T.~N.~Thorpe, J.~Yamaoka and J.~A.~Kadyk {\it et al.},
  ``3-D Tracking of Nuclear Recoils in a Miniature Time Projection Chamber,''
  arXiv:1407.7013 [physics.ins-det].
  %%CITATION = ARXIV:1407.7013;%%

\bibitem{magboltz}
All drift velocities used were calculated with the CERN-produced
software Magboltz, version 10.0.1,
10.\\http://consult.cern.ch/writeup/magboltz/

\bibitem{beasttpc}
 I.~Jaegle, S.E.~Vahsen, {\it et al.},
 ``High Resolution 3-D Tracking in a TPC with Pixel Readout,
  in preparation (2014).

\bibitem{Sauli:1997qp}
  F.~Sauli,
  ``GEM: A new concept for electron amplification in gas detectors,''
  Nucl.\ Instrum.\ Meth.\  {\bf A386}, 531-534 (1997).

\bibitem{Aad:2008zz}
  G.~Aad, M.~Ackers, F.~A.~Alberti, M.~Aleppo, G.~Alimonti, J.~Alonso, E.~C.~Anderssen, A.~Andreani {\it et al.},
  ``ATLAS pixel detector electronics and sensors,''
  JINST {\bf 3}, P07007 (2008).

\bibitem{kadyk}
J.A.~Kadyk, {\it et al.},``Conductive Coatings on a Pixel Chip Detector'',
in preparation (2014).



\end{thebibliography}

\end{document}